\documentclass{iau}
\usepackage{graphicx,amsmath}

\title[Spots, flares, accretion and obscuration in DQ Tau] 
      {Spots, flares, accretion, and obscuration in the pre-main sequence binary DQ Tau}

\author[\'A.~K\'osp\'al et al.]  
       {\'A. K\'osp\'al$^{1,2}$, P. \'Abrah\'am$^1$,
         G. Zsidi$^1$, K. Vida$^1$, R. Szab\'o$^1$,
         A. Mo\'or$^1$, and A. P\'al$^{1}$}

\affiliation{$^1$Konkoly Observatory, Research Centre for Astronomy
  and Earth Sciences, Hungarian Academy of Sciences, Konkoly-Thege
  Mikl\'os \'ut 15-17, H-1121, Budapest, Hungary \\[\affilskip] $^2$Max Planck Institute for
  Astronomy, K\"onigstuhl 17, D-69117 Heidelberg, Germany}

\pubyear{2019}
\volume{345}  
\jname{Origins: from the Protosun to the First Steps of Life}
\editors{Bruce G. Elmegreen, L. Viktor T\'oth, Manuel G\"udel, eds.}

\begin{document}

\maketitle

\begin{abstract} DQ~Tau is a young low-mass spectroscopic binary,
  consisting of two almost equal-mass stars on a 15.8\,day period
  surrounded by a circumbinary disk. We analyzed DQ~Tau's light curves
  obtained by Kepler K2, the Spitzer Space Telescope, and ground-based
  facilities. We observed variability phenomena, including rotational
  modulation by stellar spots, energetic stellar flares, brightening
  events around periastron due to increased accretion, and short dips
  due to temporary circumstellar obscuration. The study on DQ~Tau will
  help in discovering and understanding the formation and evolution of
  other real-world examples of ``Tatooine-like'' systems. This is
  especially important because more and more evidence points to the
  possibility that all Sun-like stars were born in binary or multiple
  systems that broke up later due to dynamical interactions.
\end{abstract}


Pre-main sequence stars are strongly linked with their circumstellar
environment, causing variability at a wide range of wavelengths and
timescales. The four main origins of photometric variability are
variable accretion, rotational modulation due to hot or cold stellar
spots, variable line-of-sight extinction, and stellar flares. The
second mission of the Kepler spacecraft (K2) provided high precision
photometry for many young stars in the Taurus star forming region in
2017 (Campaign 13), enabling us to search for young stellar
variability, disentangle the different effects, and determine
their relative contributions.

DQ~Tau consists of two almost identical 0.6$\,M_{\odot}$ stars that
orbit each other in an eccentric orbit with a period of 15.8\,days. At
periastron they approach each other within 12.5$\,R_{\odot}$. The
binary is surrounded by a circumbinary protoplanetary disk, from which
gas and dust are accreting onto the stars. The system is seen close to
pole-on (\cite{czekala2016}), it displays quasi-periodic optical
variability due to pulsed accretion (\cite{tofflemire2017}), and
millimeter flares and elevated X-ray activity near periastron due to a
combination of magnetic and dynamic effects (\cite{getman2011,
  kospal2011}).

\begin{figure}[b]
\vspace*{-.1 cm}
\begin{center}
 \includegraphics[width=5.3in]{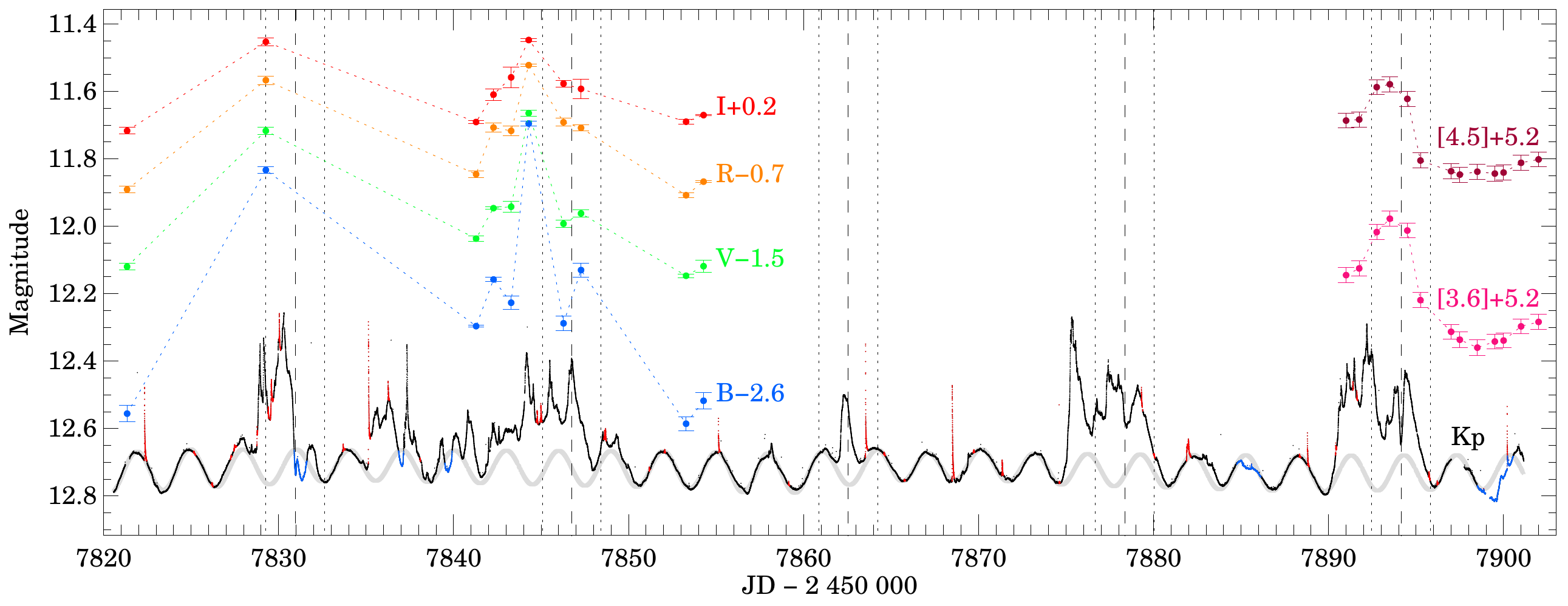}
\vspace*{-.1 cm}
 \caption{Light curves of DQ~Tau at different wavelengths (\cite{kospal2018}).}
   \label{fig:light}
\end{center}
\end{figure}

Kepler K2 observed DQ~Tau in 2017, providing an 80-day-long
uninterrupted monitoring of the white-light brightness with a cadence
of 1\,min (Fig.~\ref{fig:light}). For a few weeks at the beginning of
the K2 monitoring, we could observe DQ~Tau from the ground using the
90\,cm Schmidt telescope of Konkoly Observatory (Hungary), with
BV(RI)$_{\rm C}$ filters. We complemented these with mid-infrared
photometry at 3.6 and 4.5$\,\mu$m using the Spitzer Space
Telescope
during the final 11 days of the K2 campaign with a cadence of
20 hours.

The K2 data displays strong periodicity with $P =
3.017\pm0.004$\,days, consistently with the stellar rotational
period. The phase-folded multi-filter optical data could be fitted
with a model containing three spots that are 400\,K cooler than the
stellar photosphere and together cover 50\% of the stellar surface
(assuming that only one star is spotted).

After subtracting the rotational modulation (gray curve in
Fig.~\ref{fig:light}), we searched for short, flare-like brightening
events (fast rise and exponential decay). We identified 40 such events
(highlighted with red in Fig.~\ref{fig:light}), which last for
100--200\,minutes. The energy released in the flares is between
4.4$\times$10$^{32}$ -- 1.2$\times$10$^{35}$\,erg, more powerful than
usual for main sequence late-type stars, but typical for young low
mass stars. The energy distribution of the flares
suggests non-thermal emission mechanism. The flares occur randomly; we
saw no correlation either with the rotational period (probably a
consequence of very extended active regions that are always visible),
or with the binary's orbital period (suggesting that the flares are
single-star flares happening just above the stellar surface rather
than between the two stars where their magnetospheres reconnect and
separate).

After subtracting the rotational modulation and the flares, the
remaining variability mostly consists of complex brightening events
clustered around the periastrons, probably caused by increased
accretion rate. This can be explained by pulsed accretion: the stars
gravitationally perturb the inner edge of the circumbinary disk during
each apoapsis passage and pull some material from the disk that
eventually lands on the binary components. The strongest peaks reach
5$\times$10$^{-8}\,M_{\odot}$/yr, when the accretion luminosity goes
up to 0.8$\,L_{\odot}$. The total mass accreted in each orbital cycle
is typically about 1.2$\times$10$^{-10}\,M_{\odot}$. Our Spitzer
monitoring revealed that the increased accretion luminosity during the
periastron events warmed up the inner disk temporarily by about
100\,K.

DQ~Tau shows short dips of $<$0.1\,mag in its light curve (highlighted
with blue in Fig.~\ref{fig:light}), reminiscent of the well-known
``dipper phenomenon'' observed in many low-mass young stars, probably
caused by dusty material lifted up from the inner edge of the disk. A
rather small amount of mass (7$\times$10$^{-15}\,M_{\odot}$) is enough
to explain the dips in DQ~Tau.

This project has received funding from the European Research Council
(ERC) under the European Union's Horizon 2020 research and innovation
programme under grant agreement No 716155 (SACCRED, PI:
\'A. K\'osp\'al). The authors acknowledge the Hungarian National
Research, Development and Innovation Office grants OTKA K-109276, OTKA
K-113117, NKFIH K-115709, as well as Lend\"ulet LP2018-7/2018. KV
is supported by the Bolyai J\'anos Research Scholarship of the
Hungarian Academy of Sciences.

\end{document}